# The Gender Balance of the Australian Space Research Community: A Snapshot From The 15th ASRC, 2015

Jonathan Horner[1,2], Alice Gorman[3], Ann Cairns[4,5,6] and Wayne Short[6]

[1] Computational Engineering and Science Research Centre, University of Southern Queensland, Toowoomba, Queensland 4350, Australia
[2] Australian Centre for Astrobiology, UNSW Australia, Sydney, New South Wales 2052, Australia
[3] Department of Archaeology, Flinders University, GPO Box 2100, Adelaide, South Australia 5001, Australia
[4] New South Wales Department of Education, New South Wales, Australia
[5] Division of Information Technology, Engineering and the Environment, University of South Australia, GPO Box 2471, Adelaide, South Australia, 5001, Australia
[6] National Space Society of Australia Ltd, GPO Box 7048, Sydney, New South Wales, 2001, Australia

**Summary:** In recent years, the striking gender imbalance in the physical sciences has been a topic for much debate. National bodies and professional societies in the astronomical and space sciences are now taking active steps to understand and address this imbalance. In order to begin this process in the Australian Space Research community, we must first understand the current state of play. In this work, we therefore present a short 'snapshot' of the current gender balance in our community, as observed at the 15th Australian Space Research Conference.

We find that, at this year's conference, male attendees outnumbered female attendees by a ratio of 3:1 (24% female). This gender balance was repeated in the distribution of conference talks and plenary presentations (25 and 22% female, respectively). Of the thirteen posters presented at the conference, twelve were presented by men (92%), a pattern repeated in the awards for the best student presentations (seven male recipients vs one female). The program and organising committees for the meeting fairly represented the gender balance of the conference attendees (28% and 30% female, respectively). These figures provide a baseline for monitoring future progress in increasing the participation of women in the field. They also suggest that the real barrier is not speaking, but in enabling conference attendance and retaining female scientists through their careers – in other words, addressing and repairing the 'leaky pipeline'.

## Introduction

In the past few years, the pervasive imbalance between the number of men and women studying and working in STEM (Science, Technology, Engineering and Maths) fields has become the subject of wide debate. Traditionally, women were supposed to be more suited to 'soft' disciplines such as biological and social sciences. Whether this was argued to be a result of differing interests or ability, it amounted to the same thing: a 'chilly climate' for women who did choose to enter STEM fields. Women have struggled to gain employment, promotion and pay at equal rates to their male counterparts. Researchers such as Cordelia Fine ([1]) have effectively debunked studies purporting to characterise female cognitive development as inimical to mathematical or physical reasoning. Nonetheless, beliefs about the inferior intellectual capacity of women abound in both unconscious and explicitly expressed forms, as described below [2]:

> Two stereotypes are prevalent: girls are not as good as boys in math, and scientific work is better suited to boys and men. As early as elementary school, children are aware of these stereotypes and can express stereotypical beliefs about which science courses are suitable for females and males.

Increasingly, STEM communities are refusing to accept these persistent stereotypes as 'natural', and are actively working to redress past discrimination. As a result, a number of initiatives have been launched to attempt to build equity in those subjects from kindergarten through to senior levels in academia and industry. Notable among these is the Athena SWAN charter in the UK, which was launched in 2005 to *'encourage and recognise commitment to advancing the careers of women in science, technology, engineering, maths and medicine (STEMM) employment in higher education and research'*[1] [3].

So successful has the Athena SWAN Charter been that Australian science organisations have initiated their own schemes, based on the Athena SWAN model. The Australian Academy of Science launched its pilot of the 'Science in Australia Gender Equity' scheme, SAGE, in 2015 ([4]). At the time of writing, SAGE has already been joined by 32 Australian institutions, including more than half of all Australia's Universities. The Women in Astronomy Chapter of the Astronomical Society of Australia has also taken a leading role. In addition to running annual workshops for the Astronomical community to build equity, in 2014 the Chapter launched the *Pleiades Awards*, directly inspired by the Athena SWAN program. The awards are given to Astronomy groups in Australia with a demonstrated commitment to equity in their workplaces [5].

One of the first steps any community can make to address questions of equity is to understand the state of play – to obtain a snapshot of the gender balance in their discipline as a starting point, and to see whether the distribution of conference presentations, organising committees, and particularly awards and plenary talks are a fair representation of the gender balance of the community. To this end, a number of studies have been carried out in Australia, the UK and the US, to determine the balance of the astronomical community in those countries. For example, in [6] the authors examine the gender balance at the UK's annual National Astronomy Meeting. In that work, they take inspiration from [7], who carried out a study of the talks at the 223rd American Astronomical Society meeting, presenting data on the gender of speakers, session chairs, and also those who asked questions.

At the 223rd AAS meeting, 225 talks were sampled, with 78 (35%) being presented by women. This was in broad agreement with the gender balance of conference attendees. One area that they did note a significant departure from the gender balance of attendees was in the gender of those asking questions at the end of talks. When the talks were chaired by a man, 80% of the questions asked came from male members of the audience, with only 20% coming from the female members. By contrast, when the session chairs were female, the gender balance of questions (66% male, 34% female) was the same as that as conference attendees.

---

[1] For more information on the Athena SWAN Charter, we direct the interested reader to http://www.ecu.ac.uk/equality-charters/athena-swan/ .

Following a similar methodology, [6] found that 28% of the attendees of the 2014 National Astronomy Meeting were female, a gender balance reflected fairly in the oral presentations and session chairs at that meeting. They found that, despite this, the same pattern emerged – women were under-represented among the questions asked after the talks, when compared to the gender balance of the conference itself.

These studies suggest that the number of female presenters at the conferences reflects the number of female delegates. The main problem is then that the number of delegates and presenters falls far below the equivalent categories for men – something that often reflects the gender balance of the broader community.

At the 14th Australian Space Research Conference in 2014, a lunchtime discussion meeting was held to examine questions of gender equity in the Australian space research community. Topics covered included the importance of mentors and role models, and cultural barriers to networking, such as venues which may not suit women (such as pubs and saunas). This was repeated at the 15th Australian Space Research Conference, at which it was suggested that the compilation of statistics on the gender balance at that conference would be a useful first step in understanding that balance in our community. In this work, we therefore present a snapshot of the gender balance at that conference, based on the attendees and conference program.

## The Gender Distribution of the 15th Australian Space Research Conference

The Australian space research community is a diverse group that intersects with a variety of fields such as astronomy, physics, engineering, geospatial sciences, education and heritage. A typical conference is attended by around 120 – 200 people, and a special effort is made to attract students through offering volunteer opportunities and student prizes. Each conference features around nine plenary talks, spread over the three days of the conference. Non-plenary sessions feature up to three parallel streams. Sessions cover propulsion systems, remote sensing, planetary science, archaeoastronomy, Indigenous sky knowledge, space archaeology, space physics, satellite systems, habitation and space medicine, space mission architecture, orbital debris, space environment, and STEM education.

Before we can begin to address any inequity in the Australian Space Research community, it is vital that we get a feel for the current state of play. As the idea to undertake this study was born of a meeting towards the end of the conference, it was not possible to survey the conference attendees in advance, or to repeat the excellent work of [6] and [7]. We therefore present a post-hoc analysis of the gender balance at the recent 15th Australian Space Research Conference.

To do this, we took the list of conference attendees and the conference program, and determined the gender of each person listed based on personal knowledge of the individuals concerned. Where an individual was not personally known to us, we searched for their details online, on their professional websites, or the homepages of their institutions or employers. We acknowledge the limitation that this might not necessarily represent the personal gender identification of individual delegates, and hope in future that such information might be obtained through an anonymised survey as part of the conference registration process.

In Table 1, we present the results obtained in this manner. In total, 191 people attended the conference, giving a total of 129 oral presentations. Across the meeting, there were nine plenary talks, and thirteen posters. We were not able to obtain a breakdown of the gender of conference attendees/presenters as a function of their career stage – this, again, is something we would look to obtain at future meetings.

*Table 1: The gender distribution across the 15$^{th}$ Australian Space Research Conference.*

|  | Male | Female | Total |
| --- | --- | --- | --- |
| Delegates | 145<br>75.9% | 46<br>24.1% | 191 |
| Talks | 89<br>74.8% | 30<br>25.2% | 129 |
| Posters | 12<br>92.3% | 1<br>7.7% | 13 |
| Plenary Presentations | 7<br>77.8% | 2<br>22.2% | 9 |
| Student Awards | 7<br>87.5% | 1<br>12.5% | 8 |
| Program Committee | 13<br>72.2% | 5<br>27.8% | 18 |
| Organising Committee | 7<br>70% | 3<br>30% | 10 |

The proportions of male and female delegates who presented a paper, plenary or poster compared to those simply attending the conference were also broadly similar. Of the male delegates, 74% presented out of the 145; the figure for women was 71%.

In Figure 1, we present the distribution of conference attendees (left) and conference speakers (right), based on the data in Table 1. It is immediately clear that the two distributions are essentially identical – with the gender distribution of speakers mirroring that of the attendees. The program and conference organising committees had essentially the same gender distribution as the conference attendees (Figure 2). The same was true of the plenary speakers at the meeting (Figure 3), with seven male speakers and two female speakers.

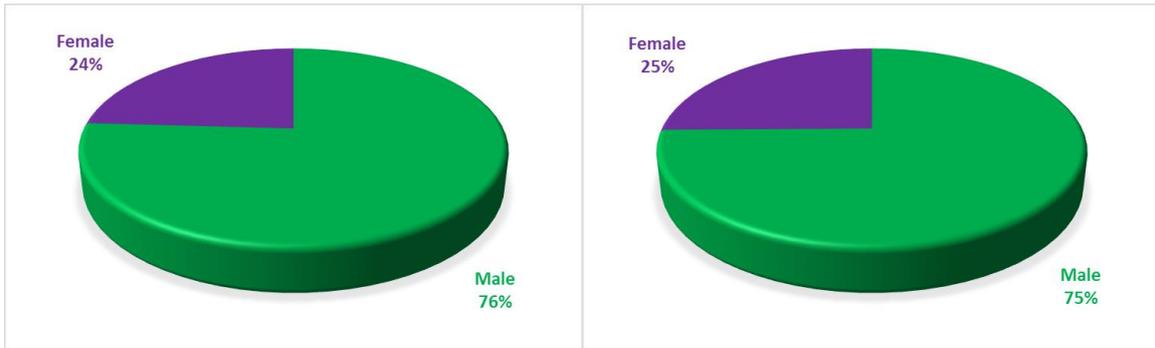

*Figure 1: The gender distribution of delegates (left; 191 total) and speakers (right; 129 total) at the 15$^{th}$ Australian Space Research Conference.*

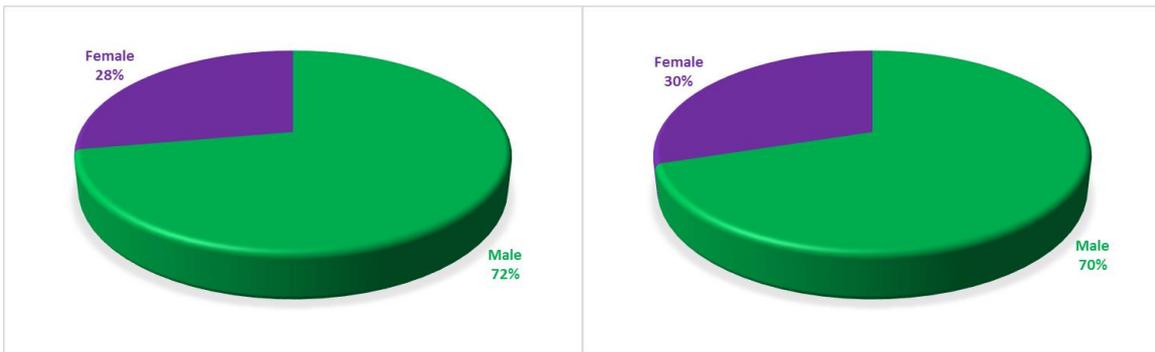

*Figure 2: The gender distribution of the program committee (left; 18 members) and organising committee (right; 10 total) at the 15$^{th}$ Australian Space Research Conference.*

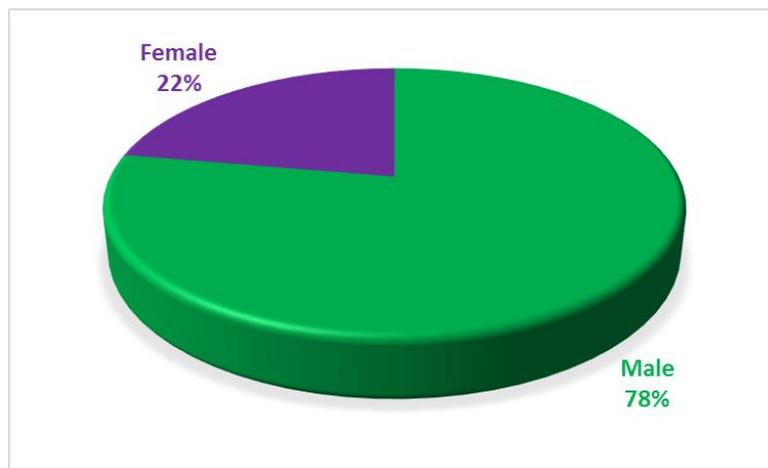

*Figure 3: The gender distribution of plenary presentations at the 15$^{th}$ Australian Space Research Conference.*

Whilst the distribution of both invited and submitted talks was in line with the overall gender distribution of conference delegates, the same is not true of the posters presented at the conference (12 male, one female) or the prizes awarded for the best undergraduate and postgraduate student presentations (seven male, one female). These figures are presented in Figure 4. Although the female percentage in both cases is lower than the overall gender balance of the conference, we note that in both cases the number of participants is so small that the difference is not strongly statistically significant.

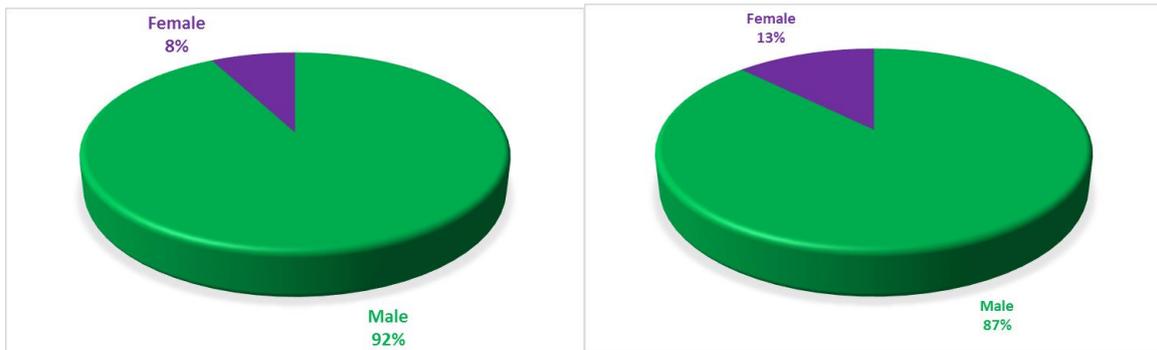

*Figure 4: The gender distribution of posters presented (left; 13 total) and prizes award (right; 8 total) at the 15th Australian Space Research Conference. The number of prizes presented to female attendees was markedly smaller than the percentage of conference attendees that were female, but not strongly statistically significant.*

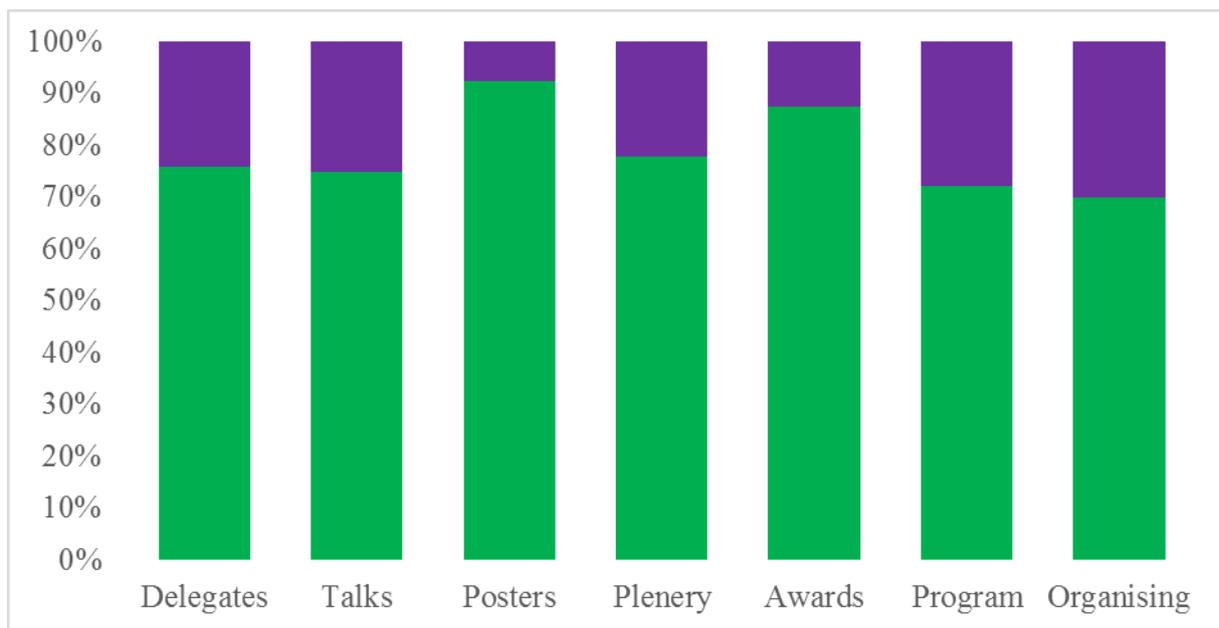

*Figure 5: The gender balance across all facets of the 15th Australian Space Research Conference. Here, green = male and purple = female. From left to right, the columns show the gender balance of: the conference delegates as a whole; the talks presented; the posters presented; the plenary presentations; the student awards; the program committee and the organising committee.*

Figure 5 presents the gender balance across the conference in the form of a bar chart, allowing the distributions in each category to be directly compared. The paucity of posters presented by women, and the male-dominated student awards clearly stand out from the overall trend of ~25% female participation at the meeting.

These results, taking into account the limitations identified, are consistent with the two studies described above. The number of female/male presenters is in direct proportion to the number of female/male delegates. This suggests that the immediate problem to be addressed is increasing the number of women attending the conference.

## Current and Future Initiatives

In order to begin to address the clear gender imbalance within the Australian Space Research Community, lunchtime discussion meetings were held at both the 14th and 15th conferences.

These meetings, which were well attended (~30 attendees in 2015) featured an open and wide-ranging discussion, covering many of the factors that are thought to contribute both to the early-career gender imbalance in our field, and the ongoing 'leaky pipeline' that has been recognised as exacerbating that imbalance towards later career stages (see e.g. Slide 6 in http://www.mso.anu.edu.au/wia-workshop-2014/presentations/WiA_workshop2014_demographics.pdf).

Several ideas have been highlighted that might help to effect long-term change, and to help make our community and conferences more equitable. These would benefit everybody – both male and female – but might help to address some of the issues that drive the 'leaky pipeline'. They include:

- The provision of childcare at the conference for delegates;
- The option to attend the conference remotely (i.e. via streaming) for those who can't attend in person;
- The organisation of networking sessions during the daytime, for those who can't attend evening events;
- Aiming to increase the fraction of female plenary speakers above parity with the current demographic distribution of attendees;
- Inviting a plenary speaker to talk on the topic of equity in space science;
- Increasing the visibility of female space scientists online by creating Wikipedia and Scimex profiles[2];
- Creating a repository of information on women working in space science (such as pictures, CVs and biographies), to provide a teaching resource.

The first four recommendations reflect the fact that women are more likely to be the primary carers of children, elderly parents or relatives, and partners, and thus less mobile, without additional support, when it comes to attending conferences.

Although no data are available to us about the gender distribution of attendees and speakers as a function of their career stage, our anecdotal observations through the course of the conference suggest that the gender balance among early-career researchers at the conference is far closer to parity. Addressing the 'leaky pipeline' to ensure that the female early-career researchers have the same opportunities for advancement as male early-career researchers is a challenge that goes beyond the remit of our conference – but taking some small measures to increase the exposure of female ECR and MCRs, and to highlight the success of strong female role models, can help to address the issue that "You can't be what you can't see".

One suggestion that was made after the conference finished was that future equity discussion meetings should possibly be embedded within the main conference program, rather than held as lunchtime splinter events, to ensure the best possible exposure and participation.

## Conclusions

In recent years, the gender imbalance in the physical sciences has become the subject of much discussion. Globally, the STEM community is taking steps to understand and address inequity in the field – and it is recognised that a vital first step in that process is to obtain and track statistics on the gender balance as a function of time, both at conferences and in the community as a whole[3].

---

[2] Scimex (https://www.scimex.org/) is a media database maintained by the Australian Science Media Centre.

For that reason, we have carried out a simple post-hoc analysis of the gender balance of the 15th Australian Space Research conference, in order to assess the current 'state of play'. Of 191 conference attendees, 46 were female (24%). Of a total of 129 oral presentations, 30 were given by women (25%). The balance of the plenary presentations (seven male, two female), and the program and organising committees (13:5 and 7:3 male:female, respectively) were all consistent with this roughly 3:1 male:female balance. By contrast, the poster presentations were strongly male dominated (twelve to one), as were the prizes awarded for the best student presentations (seven to one). This provides us with some baseline figures to use for comparison in subsequent years.

Future research may explore gendered factors in the questions addressed formally to a speaker after their presentations, both in terms of the speaker's gender, and those asking the questions, as discussed by [6] and [7]. It would also be useful to compare career stage, to investigate the observation that there was a higher proportion of ECR women at the conference. Posters are frequently a pathway for students to more fully participate, and it is noteworthy that there was only one poster by a women out of the 12 at the 2015 ASRC. As noted, this number is not statistically significant, but perhaps flags the need to obtain further data. On the other hand, this may also be due to conscious action to increase speaking opportunities for women.

Our results show that when women attend a conference such as the ASRC, they participate as actively in presenting their research as do male delegates. The problem, therefore, lies in increasing the number of women attending the conference. Recommendations arising from the 2015 meeting emphasised the necessity for childcare and other flexible attendance options. Currently these are not automatically provided as part of the conference logistics and are something for the organising and program committees to consider.

Most studies of gender balance in conferences focus on the number of female vs male speakers on panels or in invited talks (e.g. [8]). The solutions proposed are generally around identifying emerging all-male line-ups and ensuring that more women are invited before the line-ups are finalised. These discussions, however, rarely address the number of women in the general conference sessions. The ASRC currently solicits papers in a general call-out in which people self-select. It is at this stage that we need to increase the number of women submitting abstracts.

While it is not possible to solve society-wide biases that work against women in STEM with a study of this nature, the small steps taken here to understand gender inequity in the context of the Australian Space Research community offers some ways forward.

# Acknowledgements

JH is supported by USQ's Strategic Research Fund: the STARWINDS project. The authors thank Marcia Tanner of the Mars Society of Australia, who took notes at the 2014 meeting.

---

[3] See, for example, the compilation of statistics hosted on the website of the Astronomical Society of Australia's Women in Astronomy chapter, at http://asawomeninastronomy.org/statistics/

# Additional Resources

The Women In Astronomy blogspot contains a collection of excellent posts by a number of academics on a variety of topics to do with equity and gender bias. http://womeninastronomy.blogspot.com.au/

The Astronomical Society of Australia's Women in Astronomy Chapter maintains a website at http://asawomeninastronomy.org/. Of particular interest are the details of the annual Women in Astronomy Workshops, which can be found here: http://asawomeninastronomy.org/meetings/. Many of the materials from those meetings are hosted on the relevant webpages and can be freely downloaded.

The 'Project Implicit' Implicit Bias tests are a fascinating tool that highlights our implicit associations. These cover a wide variety of topics, from broad fields such as age, sexuality and religion, to more specifically focussed topics such as Gender-Science. They can be found here: https://implicit.harvard.edu/implicit/selectatest.html